\documentclass[prl,aps,amssymb,showpacs,twocolumn]{revtex4}

\usepackage{dcolumn}
\usepackage{bm}
\usepackage{graphicx}
\usepackage{amsmath}
\usepackage{amssymb}
\usepackage{amsthm}
\usepackage{amsfonts}
\usepackage{enumerate}
\usepackage{latexsym}

\input{epsf}

\begin{document}

\title{A vortex dynamics approach to the Nernst effect in fluctuating superconductors}

\author{S. Raghu$^{1}$, D. Podolsky$^2$, A. Vishwanath$^2$, and David A. Huse$^3$}
\affiliation{$^1$Department of Physics, Stanford University,
Stanford, CA 94305} \affiliation{$^2$Department of Physics,
University of California, Berkeley, CA 94720}
\affiliation{$^3$Department of Physics, Princeton University,
Princeton, NJ 08544}
\date{January 17, 2008}

\begin{abstract}
We present a new method to study the Nernst effect and diamagetism
of an extreme type-II superconductor dominated by phase
fluctuations. We work directly with vortex variables and our
method allows us to tune vortex parameters (e.g., core energy and
number of vortex species).  We find that diamagnetic response and
transverse thermoelectric conductivity ($\alpha_{xy}$) persist
well above the Kosterlitz-Thouless transition temperature, and
become more pronounced as the vortex core energy is increased.
However, they \textit{weaken} as the number of internal vortex
states are increased.  We find that $\alpha_{xy}$ closely tracks the magnetization
$(-M/T)$ over a wide range of parameters.
\end{abstract}

\pacs{74.20.De, 74.25.Fy, 74.40+k, 74.72.-h }

\maketitle

A number of experimental observations in superfluidity and
superconductivity are best explained in terms of the statistical
mechanics and dynamics of vortices.  Examples include the
Kosterlitz-Thouless (KT) transition
 in superfluid
films, and the  ``flux-flow'' contribution to electrical
resistivity in type-II superconductors.  A natural extension of
this paradigm involves the use of thermoelectric and thermal
transport experiments as a probe of vortex dynamics.
This is especially pressing given the Nernst effect experiments in
the cuprates \cite{Wang2006}.
The Nernst effect is defined as the appearance of a steady-state
electric field ($E_y$) when a system is placed in a perpendicular
thermal gradient ($\nabla_x T$) and magnetic field ($H^z$) under
open circuit conditions. In the cuprate superconductors, a large
Nernst signal and diamagnetic response persists well above $T_c$,
the superconducting transition temperature, and are especially
enhanced in the underdoped regime of hole-doped materials.

To explain these results, it has been argued that vortices are
responsible for the large Nernst signal \cite{Wang2006}: as a
vortex drifts down the temperature gradient, it generates a
transverse electric field.  A vortex description above $T_c$ is
useful when the superconducting transition represents the loss of
macroscopic phase coherence, while the amplitude of the order
parameter remains large below a higher `mean-field' temperature
scale $T_c^{MF} \gg T_c$ \cite{Emery1995}.
This idea is most likely to hold true in the underdoped regime of
the hole-doped cuprates, where the reduced superfluid stiffness
enables phase fluctuations to suppress $T_c$ well below the
mean-field transition temperature.  This ``vortex-liquid regime''
regime $T_c <T\ll T_c^{MF}$ \cite{Fisher1991} over a range of
fields
can be intuitively understood in terms of a dilute fluid of
vortices, but so far this appealing picture
\cite{Ioffe2002,Honerkamp2004} has not lead to a quantitative
understanding of the Nernst effect.

There are special challenges involved in constructing a vortex
based theory of thermoelectric transport that is consistent with
the basic principles of statistical mechanics. In contrast to
thermodynamics, and even electrical transport, where a theory of
vortices interacting via a long range potential can be used, a
thermal or Nernst transport calculation requires a purely local
formulation where no such long range forces are explicitly
present. In this Letter, we present such a local formulation, and
use it to study thermoelectric transport directly in the vortex
language.
An advantage of our method is that vortex parameters
such as core energy, or number of vortex species, can be
tuned independently of other properties, and their impact on the
Nernst signal and magnetization can be systematically studied.
Moreover, it is possible to use our method to study the Coulomb
gas efficiently even in the high density limit, where the number
of pairwise Coulomb interactions is large.

Previous theoretical work on the Nernst effect in cuprates
\cite{Ussishkin2002,
Mukerjee2004,Anderson2006,Anderson2007,Sachdev2007} has studied
the time-dependent Ginzburg-Landau theory, 
which includes fluctuations in the amplitude of the order
parameter.  
In earlier work, three of us have studied the Nernst effect in an
XY model \cite{Podolsky2007}, which should be relevant for the
underdoped cuprates. Our present results approaching from the
vortex viewpoint are consistent with this earlier approach.
Remarkably, we find that the close quantitative connection between
the Nernst effect and diamagnetism observed in that and other
papers, holds in the present study too, even when vortex
properties are drastically modified.  The effect of vortex core energy on diamagnetism and the 
transition temperature was studied in refs. \cite{Benfatto2007A,Benfatto2007B,Benfatto2007c}.

\textit{Method:} Consider a two dimensional superconductor in the
extreme type-II limit, in which the supercurrents are too feeble
to modify the externally imposed field. Since we are interested in
the `vortex liquid' regime, we restrict the order parameter to
live on the sites $n$ of a square lattice and assume that all
fluctuations arise from the phase:
$\psi_{n}(t)=|\psi_0|e^{i\theta_{n}(t)}$
We map to the vortex representation by defining a {\it dual}
electric field $\bm e = (e^x_{i},e^y_{i})$ on the bonds
$(i,i+\hat{x});\, (i,i+\hat{y})$ of a dual lattice, orthogonal to
the original lattice bonds, via:
$\bm e = \nabla \theta \times \hat{ \bm z}$
where $\nabla$ is a lattice derivative. The local field ${\bm
e}$ is related to the local supercurrents via $ \bm J = \rho_s^0
\hat{\bm z} \times \bm e $, where $\rho_s^0$ is the bare
superfluid density. Vortices may live at the sites $i$ of this
dual lattice, and the integer vortex charge $n_i$ satisfies Gauss'
law $ \nabla\cdot {\bm e}= 2\pi n_i$.
We thus obtain a completely local Hamiltonian for the vortex fluid
that retains all phase and vortex degrees of freedom:

\begin{equation}
\mathcal{H} = \frac{1}{4 \pi} \sum_{i}   \bm e_i^2 +\epsilon_c
\sum_{i}   n_{i} ^2,
\end{equation}
where $\epsilon_c$ is the vortex core-energy and we use units
where
$\rho_s^0= 1/2 \pi$.
When the transverse electric fields are integrated out, the model
reduces to the static 2D Coulomb gas. However,  we stress that one
must include both local supercurrents and vortices when
dealing with thermal transport, so as to define a local energy
density.  The interaction between the vortices is mediated by
the supercurrents, and this maps simply to the interaction between
the charges being mediated by the dual electric field in the
Coulomb gas.

\begin{figure}
\includegraphics[width=1.5in]{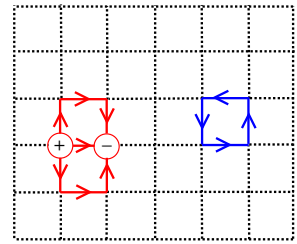}
\caption{Charge updates (left) and curl updates (right) sample the
longitudinal and transverse degrees of freedom, respectively,
of the dual electric field. } \label{mc}
\end{figure}

The model is given a Monte Carlo dynamics that captures the effect
of random thermal fluctuations.
Two distinct types of updates, charge and curl updates,
corresponding to the two degrees of freedom (vortices and supercurrents )
are introduced (Fig. \ref{mc}).
During a curl update, a plaquette is chosen at random and a random
electric curl is added to it.  Such an electric flux configuration
is purely transverse and is not accompanied by vortex creation.
During a charge update, a lattice bond
is chosen at random, and a vortex/anti-vortex pair
is added on the two sites connected to this bond. The electric
fluxes are updated locally near the charges to satisfy Gauss' law.
 Note, this charge update may result in the motion a pre-existing vortex.  In
the simulation,  each move is accepted with probability $ 1/
\left[ 1+\exp \left( \Delta U/T \right) \right] $, where $\Delta
U$ is the change in energy associated with the move and $T$
is the local temperature at the center of the plaquette or bond
for that move.
By varying the relative frequency of each type of trial, we have
control over $D_{ph}$, the phase diffusivity, relative to
the vortex diffusivity $D_v$.  In what follows, we work in the
physically reasonable limit $D_{ph} \gg D_v$. After an attempt is
made to update each bond and plaquette with such moves, a unit of
Monte Carlo time passes in the simulation.  A related method
(without pair creation) has been used in Ref.~\cite{Maggs2002,
Levrel2005} to study charged polymers.

When a charge move is attempted, the simplest way to satisfy
Gauss' law is to add an electric flux $ \bm e = 2 \pi \hat{\bm r}
$ to the link $\hat{\bm r}$ connecting the positive to the
negative charge. However, such single bond updates usually
produce large $\Delta
U$ and thus are rarely accepted, 
resulting in too little vortex motion at low temperatures; hence
an
update that spreads the electric flux over several bonds is used.
A simple example of such an update is shown in Fig.~\ref{mc}.
The added flux is made curl-free and the move we actually use
involves a patch of 12 plaquettes that is one plaquette larger in
all directions than indicated in Fig.~\ref{mc}.

{\em Thermodynamics:} We have studied the KT transition of the
neutral 2D Coulomb gas by tracking the dielectric response
function
\begin{equation}
\epsilon^{-1} = 1 - \frac{\langle \left(\sum_i \bm e_i \right)^2
\rangle }{4 \pi T L^2}. \label{inveps2}
\end{equation}
\begin{figure}
\includegraphics[width=2.8in]{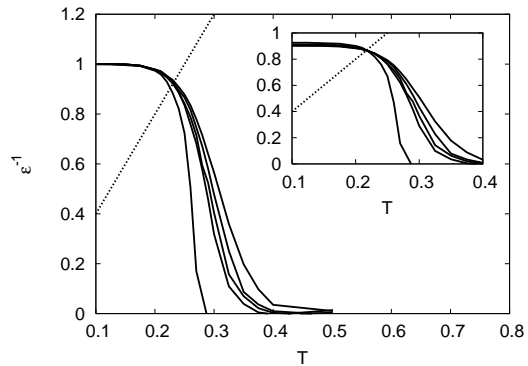}
\caption{Dielectric response of the dual Coulomb gas, computed
using Eq. \ref{inveps2}, and (inset) rescaled for finite-size
effects \cite{Weber1988}. We use an $L \times L$ torus with $ L =
8,10,12,14,40$, and core energy $\epsilon_c = 0.125$. The
transition temperature is the location where the rescaled data
intersects the line $y=4T$ (dotted line). } \label{hmod}
\end{figure}
At $T=T_{KT}$, the dielectric response function satisfies the
universal jump criterion \cite{Nelson1977}:
\begin{equation}
\epsilon^{-1} = \left\{ \begin{array}{ll}
4T_{KT} \ , &  T=T_{KT}^- \\
0 \ , & T=T_{KT}^+  \end{array} \right.
\end{equation}
Fig. \ref{hmod} shows the results of the helicity modulus
calculation using various system sizes and a core energy
$\epsilon_c = 0.125$.   We use finite-size scaling
\cite{Weber1988} to identify $T_{KT} = 0.22$; the naive
free-energy estimate is $T^0_{KT}=0.25$ in our units.
In the remainder of  this letter, we cite all temperatures in
units of $T_{KT}^0$.

\textit{Diamagnetism:} When a magnetic field is applied, the
resulting imbalance of vorticity leads us to consider a
plasma
of vortices in a static, neutralizing background (charge density
$n_b =-B/{\Phi_0}$), where $\Phi_0= 2 \pi $ is the flux
quantum in our units. 
To study diamagnetism we must permit vortices to enter and
leave the sample, so we employ a cylindrical geometry with open
boundaries.  Current flow near the boundaries is measured, which
arises due to a surface depletion of vorticity. 
This has a lower free energy than the perfectly neutral system,
and leads to diamagnetism.
Deep inside the cylinder, beyond some distance $x_0$ from the
edge, the supercurrents vanish in equilibrium.
The magnetization can be obtained by inverting the relation $\bm J
= \bm \nabla \times \bm M$ on a cylinder whose axis is along the
$x$ direction. For physical clarity we present continuum formulas
below, these can be readily transcribed into the appropriate
lattice versions. We have:
\begin{equation}
M = \int_0^{x_0} dx \ \langle J^y (x,y) \rangle = \rho_s
\int_0^{x_0}  dx  \ \langle e_x \rangle~. \label{magnetization}
\end{equation}
Thus, the magnetization is directly proportional to the
\textit{work function} of the dual Coulomb gas (energy cost of
removing a vortex from the bulk of the system).
Using Gauss' law we can also obtain:

\begin{equation}
M = 2 \pi \rho_s \int_0^{x_0}   dx  \ x  \ \left( \langle n(x)
\rangle - B/\Phi_0 \right).
\end{equation}
Thus, the magnetization is also the total edge polarization
(dipole moment) per unit length. 
The polarization fields are non-zero only in the charge depletion
region near the cylinder's edge.

{\em Nernst Effect:} To determine the Nernst effect, we again make
use of cylindrical geometry, and apply a temperature gradient
along the cylinder axis.
In our simulations, we compute the transverse thermoelectric
conductivity $\alpha_{xy}$ defined via $\langle J_y \rangle =
-\alpha_{xy} ( - \nabla_x T)$.  The thermoelectric conductivity is
closely related to the Nernst signal: $\alpha_{xy}/\sigma_{xx}$,
where $\sigma_{xx}$ is the electrical conductivity and we have
made the approximation of vanishing Hall angle. In our
calculations, $\alpha_{xy}$ is obtained by measuring the dual
electric field $e^x$, and using the relation
$\alpha_{xy} = \langle e^x \rangle /\nabla_x T$. Thus while
$\alpha_{xy}$ is an off-diagonal transport coefficient when
written in terms of electrical currents, it is a \textit{diagonal}
response function in the vortex representation: it is simply the
vortex {\em thermopower}.  We have verified that the net vortex
motion vanishes once the steady-state in the thermal gradient is
reached, as is required in a thermopower measurement.  The
quantity $\alpha_{xy}$ has the advantage that, unlike the Nernst
signal $e_N$, it does not have any explicit dependence on
$t_{mc}$, the Monte-Carlo time step, as can be seen from the Kubo
formula for $\alpha_{xy}$ and from dimensional analysis
\cite{Podolsky2007}.

\begin{figure}
\includegraphics[width=3.5in]{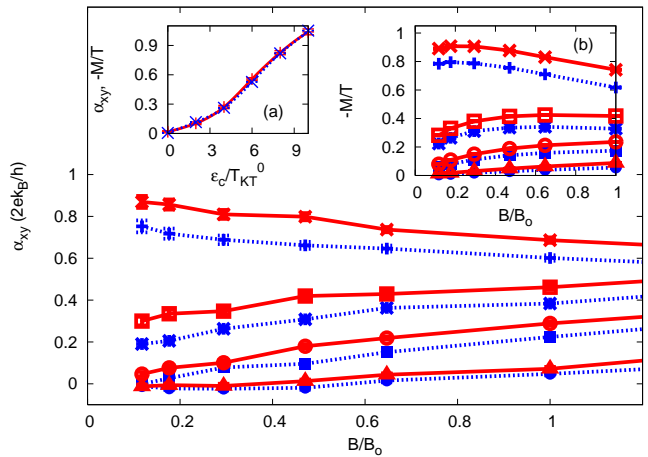}
\caption{ $\alpha_{xy}$ for a system of single-flavor (solid
lines) and 2-flavor (dotted lines) vortices on a 20x15 cylinder,
 with core energy $\epsilon_c=0.5T_{KT}^0$. Temperatures shown are
$1.0$ (topmost 2 curves), $1.25, 1.5$, and $2.0$ $T_{KT}^0$
(bottom 2 curves). \emph{Inset a}: Core energy dependence of
$\alpha_{xy}$ (solid lines) and $-M/T$ (dashed lines)  in
units of $2ek_B/h$ for single-flavor vortices, at
$T=2.5T_{KT}^0$, $B=0.7B_0$ showing a marked increase of both
quantities with core energy. \emph{Inset b}: Diamagnetism $-M/T$
curves for the same temperatures as the main figure. (Units of
$2ek_B/h$ are used on the vertical axes of both insets.) }
\label{axy}
\end{figure}

Figure \ref{axy} shows the simulation results for $\alpha_{xy}$
and $M$ (inset b)
 for applied fields up
to $B_0 = \Phi_0/(2 \pi a^2)$, where $a$ is the lattice spacing,
comparable to the zero temperature coherence length $\xi_0$. Both
$\alpha_{xy}$ and $-M/T$ are expressed in units of the 2D
``quantum of thermoelectric conductance," $2ek_B/h$.  We show data
for $T \geq T_{KT}^0$; below $T_{KT}^0$, our simulations encounter
difficulties due to impaired vortex mobility, and furthermore, it
is difficult to apply thermal gradients small enough to remain
within linear response.  Fortunately, $T \geq T_{KT}^0$ is a regime
of interest, since the Nernst signal persists well above $T_c$ in
the experiments of Wang, \emph{et al.} \cite{Wang2006}. The
results presented here are in quantitative agreement with earlier
computations involving the 2D XY model with Langevin dynamics
\cite{Podolsky2007}: in particular, both models have the feature
that in the small magnetic field limit, $a)$ $\alpha_{xy}$ and $M$
diverge logarithmically as $B \rightarrow 0$ for $T \leq T_{KT}$
and $b)$ they increase linearly with $B$ at small $B$ when
$T>T_{KT}$ \cite{Oganesyan2006}.  Similar features have been seen
in experiments of Wang, \emph{et al.} \cite{Wang2006}. Moreover,
both $\alpha_{xy}$ and $M$ are detectable at temperatures as high
as $2T_{KT}$ in this vortex model, which is also consistent with
2D XY model results \cite{Podolsky2007}.  At very high
temperatures $T \gg T_{KT}$, the $2D$ XY model predicts that
$\alpha_{xy}$ and $M/T$ decay sharply as a power law in
temperature. In the vortex model, however, the magnetization
decreases even more rapidly at such high temperatures: a
calculation based on the dual solid-on-solid model, shows that the
magnetization of single-flavor vortices decays exponentially as:
$M = -2T \sin \left( B/B_0 \right)e^{-2T/ \rho_s}$
We have checked that our numerical results agree
with this expression. Although we have not succeeded in finding
similar expressions for $\alpha_{xy}$, our numerical results
indicate that $\alpha_{xy}$ also decays in this fashion at high
temperatures and closely tracks the diamagnetism.

{\em Vortex Core Energy Dependence:} The core energy
dependence of $\alpha_{xy}$ and diamagnetism $-M/T$ are shown in
Fig. \ref{axy} (inset a) at $T=2.5T_{KT}^0$.  So long as
$\epsilon_c \not\gg T$, both are found to increase with
$\epsilon_c$.  At this
temperature, $\alpha_{xy}$ and $-M/T$ 
track each other closely.  
With increasing core energy, $\alpha_{xy}$ and $-M/T$ rise from
near zero at $\epsilon_c = 0$ to $O(1)$ at $\epsilon_c = 10
T_{KT}^0$, showing that the core energy has a dramatic impact on
both of these quantities in this regime.  The dominant effect of
the vortex core energy $\epsilon_c$ is that it enhances local
superconducting correlations at short distances by increasing the
cost of vortex fluctuations.  The core energy enters directly in
setting the ``work function'' for removing a vortex from the
system, which is proportional to the magnetization.  For
$\epsilon_c \gg T$, we thus expect $-M \propto \epsilon_c$, since
more vortices are expelled near the boundaries in this limit.   We
have observed this in our simulations.  The vortex-free boundary
layer grows with $\epsilon_c$ and when it becomes comparable to
the thickness of the sample this causes strong finite-size
effects. Thus we have not been able to reliably determine the bulk
behavior for large values of the vortex core energy $\epsilon_c$.
We expect $\alpha_{xy}$ to saturate at a finite value in this
limit -- when all thermally-generated vortex fluctuations are
suppressed, the remaining field-induced vortices respond to the
thermal gradient in a way that is independent of the magnitude of
$\epsilon_c$.   Although we do see some indications of the saturation of
$\alpha_{xy}$ at large values of $\epsilon_c$, we have
not been able to access this regime reliably due to the large
finite-size effects on $\alpha_{xy}$ when $\epsilon_c
>> T$.

{\em Dependence on the Number of Vortex Flavors:} Several theories
of cuprates predict additional degrees of freedom associated with
vortices, that endows them with a flavor index
\cite{Lee2001,Arovas1997}. Such an extension is readily
incorporated in our formalism, the Gauss law is now $ \nabla\cdot
{\bm e}= 2\pi \sum_{\alpha=1}^{N_v}n_{i\alpha}$ where $\alpha$ is
the flavor index .
Our data shows a systematic dependence on the number of internal
flavors of vortices. We find that both $\vert M \vert$ and
$\alpha_{xy}$ of single-flavor vortices (solid curves in
Fig.~\ref{axy})  are systematically larger than that of vortices
with 2 internal flavors (dotted lines).  With multiple vortex
species, there is additional configurational entropy associated
with the internal degree of freedom.  Therefore, the free energy
cost of introducing a vortex into the sample is lower, and the
system becomes less diamagnetic. It is more difficult to
understand, however, why $\alpha_{xy}$ decreases as the number of
species is increased.  In phenomenological discussions, the
transport coefficient $\alpha_{xy}$ is usually identified with the vortex
entropy \cite{Caroli1967}.
Therefore, one may naively expect $\alpha_{xy}$ to increase with
the number of internal vortex flavors. Moreover, $\alpha_{xy}$ is
equivalent to the vortex thermopower, and if we neglected the
logarithmic interactions between vortices, a simple calculation
shows that the thermopower of an ideal gas of vortices
is simply the entropy per particle \cite{Raghu2007};
which increases linearly with the number of internal states
associated with such particles. Our numerical results however are
in sharp contrast to such expectations that $\alpha_{xy}$ ought to
be proportional to an effective entropy per vortex. Instead, our
results which have the opposite trend point towards a scenario
that is far more complicated than an ideal gas approximation;
interactions between vortices and thermally generated
vortex-antivortex pairs invalidate such simple relations between
$\alpha_{xy}$ and vortex entropy.

Perhaps the most striking feature of our data is the fact that
$\alpha_{xy}$ and $-M/T$ closely track each other.  In particular,
at high temperatures, our results obey the relation $\alpha_{xy}
= -cM/T$, where $c \approx 1$, and is difficult to determine
accurately due to noise at large temperatures.  Recently, it
was shown  analytically in Ref. \cite{Podolsky2007}  that for the
2D XY Hamiltonian with overdamped dynamics, $c = 1/2$ at high
temperatures.  Furthermore, one obtains the same relation for
Gaussian superconducting fluctuations \cite{Ussishkin}.  Here, we
have shown that this relation is even more robust, since it hardly
depends on the variation of the core energy and the number of
distinct vortex species so long as $\epsilon_c$ is not much
greater than $T$.  This relationship which connects  a transport
coefficient and a thermodynamic quantity is indicative of a deep
underlying (and as yet unknown) principle, that seems to hold true
over a broad range of parameters.

In conclusion, we have presented a new local method to study the
thermodynamic and transport properties of 2D vortex liquids.  This
method enables us to directly observe the effect of tuning vortex
parameters on $\alpha_{xy}$ and $M$.  As an application of our
method, we have systematically studied the effect of modifying the
vortex core energy and the number of vortex species on
diamagnetism and Nernst effect. In all cases, we have found that
both quantities persist well above $T_{KT}$ and $\alpha_{xy}$
closely tracks $-M/T$ when $T_{KT} < T \not\ll \epsilon_c$.
We have provided the first detailed analysis of the Nernst effect
of a vortex liquid that deals directly with vortex variables. The
method presented here can be generalized to study thermal
transport of vortex liquids in 3D, and can be extended to study
quantum phase fluctuations at low temperatures.

We thank P. W. Anderson, P.A. Lee and N. P. Ong for very helpful
discussions. This work was supported by NSF DMR-0645691 (A. V.),
-0213706 (D.A.H.) and by the Stanford Institute for Theoretical
Physics (S. R).


\end{document}